\documentclass{ws-procs975x65}

\begin{document}

\title{Lattice Gauge Theory and (Quasi)-Conformal Technicolor}

\author{D. K. SINCLAIR}

\address{HEP Division, Argonne National Laboratory, 9700 South Cass Avenue, \\
Argonne, Illinois 60439, USA \\
$^*$E-mail: dks@hep.anl.gov}

\author{J. B. KOGUT}

\address{Department of Energy, Division of High Energy Physics, \\
Washington,DC 20585, USA \\
and \\
Department of Physics -- TQHN, University of Maryland, 82 Regents Drive, \\
College Park, Maryland 20742, USA \\
E-mail: jbkogut@umd.edu}

\begin{abstract}
QCD with 2 flavours of massless colour-sextet quarks is studied as a theory
which might exhibit a range of scales over which the running coupling constant
evolves very slowly (walks). We simulate lattice QCD with 2 flavours of sextet
staggered quarks to determine whether walks, or if it has an infrared fixed
point, making it a conformal field theory. Our initial simulations are
performed at finite temperatures $T=1/N_ta$ ($N_t=4$ and $N_t=6$), which
allows us to identify the scales of confinement and chiral-symmetry breaking
from the deconfinement and chiral-symmetry restoring transitions. Unlike QCD
with fundamental quarks, these two transitions appear to be well-separated.
The change in coupling constants at these transitions between the two
different temporal extents $N_t$, is consistent with these being finite
temperature transitions for an asymptotically free theory, which favours
walking behaviour. In the deconfined phase, the Wilson Line shows a 3-state
signal. Between the confinement and chiral transitions, there is an additional
transition where the states with Wilson Lines oriented in the directions of
the complex cube roots of unity disorder into a state with a negative Wilson
Line.
\end{abstract}


\bodymatter

\chardef\quoteleftcode=\catcode96	
\catcode96=12				

\newread\epsffilein    
\newif\ifepsffileok    
\newif\ifepsfbbfound   
\newif\ifepsfverbose   
\newdimen\epsfxsize    
\newdimen\epsfysize    
\newdimen\epsftsize    
\newdimen\epsfrsize    
\newdimen\epsftmp      
\newdimen\pspoints     
\pspoints=1bp	       
\epsfxsize=0pt	       
\epsfysize=0pt	       
\def\epsfbox#1{\global\def\epsfllx{72}\global\def\epsflly{72}%
   \global\def\epsfurx{540}\global\def\epsfury{720}%
   \def\lbracket{[}\def\testit{#1}\ifx\testit\lbracket
   \let\next=\epsfgetlitbb\else\let\next=\epsfnormal\fi\next{#1}}%
\def\epsfgetlitbb#1#2 #3 #4 #5]#6{\epsfgrab #2 #3 #4 #5 .\\%
   \epsfsetgraph{#6}}%
\def\epsfnormal#1{\epsfgetbb{#1}\epsfsetgraph{#1}}%
\def\epsfgetbb#1{%
%
%
\openin\epsffilein=#1
\ifeof\epsffilein\errmessage{I couldn't open #1, will ignore it}\else
%
%
   {\epsffileoktrue \chardef\other=12
    \def\do##1{\catcode`##1=\other}\dospecials \catcode`\ =10
    \catcode`\^^L=9 \catcode`\^^?=9
    \loop
       \read\epsffilein to \epsffileline
       \ifeof\epsffilein\epsffileokfalse\else
%
%
	  \expandafter\epsfaux\epsffileline:. \\%
       \fi
   \ifepsffileok\repeat
   \ifepsfbbfound\else
    \ifepsfverbose\message{No bounding box comment in #1; using defaults}\fi\fi
   }\closein\epsffilein\fi}%
%
%
\def\epsfclipstring{}
\def\epsfclipon{\def\epsfclipstring{ clip}}%
\def\epsfclipoff{\def\epsfclipstring{}}%
\def\epsfsetgraph#1{%
   \epsfrsize=\epsfury\pspoints
   \advance\epsfrsize by-\epsflly\pspoints
   \epsftsize=\epsfurx\pspoints
   \advance\epsftsize by-\epsfllx\pspoints
%
%
   \epsfxsize\epsfsize\epsftsize\epsfrsize
   \ifnum\epsfxsize=0 \ifnum\epsfysize=0
      \epsfxsize=\epsftsize \epsfysize=\epsfrsize
      \epsfrsize=0pt
%
%
     \else\epsftmp=\epsftsize \divide\epsftmp\epsfrsize
       \epsfxsize=\epsfysize \multiply\epsfxsize\epsftmp
       \multiply\epsftmp\epsfrsize \advance\epsftsize-\epsftmp
       \epsftmp=\epsfysize
       \loop \advance\epsftsize\epsftsize \divide\epsftmp 2
       \ifnum\epsftmp>0
          \ifnum\epsftsize<\epsfrsize\else
	     \advance\epsftsize-\epsfrsize \advance\epsfxsize\epsftmp \fi
       \repeat
       \epsfrsize=0pt
     \fi
   \else \ifnum\epsfysize=0
     \epsftmp=\epsfrsize \divide\epsftmp\epsftsize
     \epsfysize=\epsfxsize \multiply\epsfysize\epsftmp
     \multiply\epsftmp\epsftsize \advance\epsfrsize-\epsftmp
     \epsftmp=\epsfxsize
     \loop \advance\epsfrsize\epsfrsize \divide\epsftmp 2
     \ifnum\epsftmp>0
	\ifnum\epsfrsize<\epsftsize\else
	   \advance\epsfrsize-\epsftsize \advance\epsfysize\epsftmp \fi
     \repeat
     \epsfrsize=0pt
    \else
     \epsfrsize=\epsfysize
    \fi
   \fi
%
%
   \ifepsfverbose\message{#1: width=\the\epsfxsize, height=\the\epsfysize}\fi
   \epsftmp=10\epsfxsize \divide\epsftmp\pspoints
   \vbox to\epsfysize{\vfil\hbox to\epsfxsize{%
      \ifnum\epsfrsize=0\relax
        \includegraphics{#1}%
      \else
        \epsfrsize=10\epsfysize \divide\epsfrsize\pspoints
        \includegraphics{#1}%
      \fi
      \hfil}}%
\global\epsfxsize=0pt\global\epsfysize=0pt}%
%
%
{\catcode`\%=12 \global\let\epsfpercent=
%
%
\long\def\epsfaux#1#2:#3\\{\ifx#1\epsfpercent
   \def\testit{#2}\ifx\testit\epsfbblit
      \epsfgrab #3 . . . \\%
      \epsffileokfalse
      \global\epsfbbfoundtrue
   \fi\else\ifx#1\par\else\fi\fi}%
%
%
\def\epsfempty{}%
\def\epsfgrab #1 #2 #3 #4 #5\\{%
\global\def\epsfllx{#1}\ifx\epsfllx\epsfempty
      \epsfgrab #2 #3 #4 #5 .\\\else
   \global\def\epsflly{#2}%
   \global\def\epsfurx{#3}\global\def\epsfury{#4}\fi}%
%
%
\def\epsfsize#1#2{\epsfxsize}
%
%
\let\epsffile=\epsfbox

\catcode`\`=\quoteleftcode		

\section{Introduction}

Technicolor theories are QCD-like gauge theories with massless fermions, whose
pion-like excitations play the role of the Higgs field in giving masses to the
$W$ and $Z$ \cite{Weinberg:1979bn,Susskind:1978ms}. 
We search for Yang-Mills gauge theories whose fermion 
content is such that the running coupling constant evolves very slowly -- walks.
Such theories can avoid the phenomenological problems which plague other
(extended-)Technicolor theories
\cite{Holdom:1981rm,Yamawaki:1985zg,Akiba:1985rr,Appelquist:1986an}.

While many studies have used fermions in the fundamental representation, with
large numbers of flavours
\cite{Kogut:1985pp,Fukugita:1987mb,Ohta:1991zi,Kim:1992pk,Brown:1992fz,
Iwasaki:1991mr,Iwasaki:2003de,Deuzeman:2008sc,Deuzeman:2009mh,
Appelquist:2009ty,Appelquist:2007hu,Jin:2008rc,Jin:2009mc,Fodor:2009wk,
Fodor:2009ff,Yamada:2009nt}, we are concentrating on higher representations of
the colour group (in particular the symmetric tensor), where 
conformality/walking can be achieved at much lower $N_f$. There have been
some studies with $SU(2)$ colour with two adjoint (symmetric tensor) fermions
\cite{Catterall:2007yx,Catterall:2008qk,Catterall:2009sb,DelDebbio:2008zf,
DelDebbio:2009fd,Bursa:2009we,Hietanen:2008mr,Hietanen:2009az}.
We are considering QCD ($SU(3)$ colour) with colour-sextet (symmetric tensor) 
quarks.

The 2-loop $\beta$-function for QCD with $N_f$ massless flavours of 
colour-sextet quarks, suggests that for 
$1\frac{28}{125} \le N_f < 3\frac{3}{10}$, either this theory will have an
infrared-stable fixed point, or a chiral condensate will form and this fixed
point will be avoided. In the first case the theory will be conformal; in the
second case it will walk. For $N_f=3$ conformal behaviour is expected. $N_f=2$
could, a priori, exhibit either behaviour. Because the quadratic Casimir
operator for sextet quarks is $2\frac{1}{2}$ times that for fundamental
quarks, it is easier for them to form a chiral condensate. 

Lattice QCD gives us a direct method to determine which option the $N_f=2$
theory chooses. We are studying the $N_f=2$ theory using staggered fermions. We
are currently performing simulations at finite temperature ($T$). Finite
temperature enables us to study the scales of confinement and chiral symmetry
breaking, and yields information on the running of the coupling constant.

Simulations using Wilson fermions by DeGrand, Shamir and Svetitsky suggest
that this theory is conformal \cite{Shamir:2008pb,DeGrand:2009hu}.
 Our simulations suggest that it walks.
The scales of confinement and chiral symmetry breaking appear to be very 
different. (This also contrasts with what was reported by DeGrand, Shamir and 
Svetitsky for simulations using Wilson quarks \cite{DeGrand:2008kx}.) Hence 
the phenomenology is
expected to be different from that of QCD with fundamental quarks and
$N_f$ in the walking window, where these two scales appear to be the same.
Preliminary studies of this theory using domain-wall quarks have been reported
in \cite{Fodor:2008hm}.

In the deconfined phase we observe states where the phase of the Wilson Line
is $\pm 2\pi/3$, and at weaker couplings, $\pi$ in addition to the expected
states with positive Wilson Lines. These have since been predicted and
observed by Machtey and Svetitsky using Wilson quarks \cite{Machtey:2009wu}.

\section{Simulations and Results}

We use the standard Wilson (triplet) plaquette action for the gauge fields
and an unimproved staggered-fermion action for the quarks. The only new feature
is that the quark fields are six-vectors in colour space and the gauge fields
on the links of the quark action are in the sextet representation of colour.
The RHMC algorithm is used to tune the number of flavours, $N_f$, to 2.

We run on $8^3 \times 4$, $12^3 \times 4$ and $12^3 \times 6$ lattices at
quark masses $m=0.005$, $m=0.01$ and $m=0.02$ in lattice units to allow 
extrapolation to the chiral ($m=0$) limit. $\beta=6/g^2$ is varied over a
range of values large enough to include the deconfinement and chiral 
transitions. Run lengths of 10,000--200,000 trajectories per $(m,\beta)$ are
used.

More details of the results presented here are given in 
ref.~\cite{Kogut:2010cz}. 

\begin{figure}[htb]
\epsfxsize=0pt
\epsffile{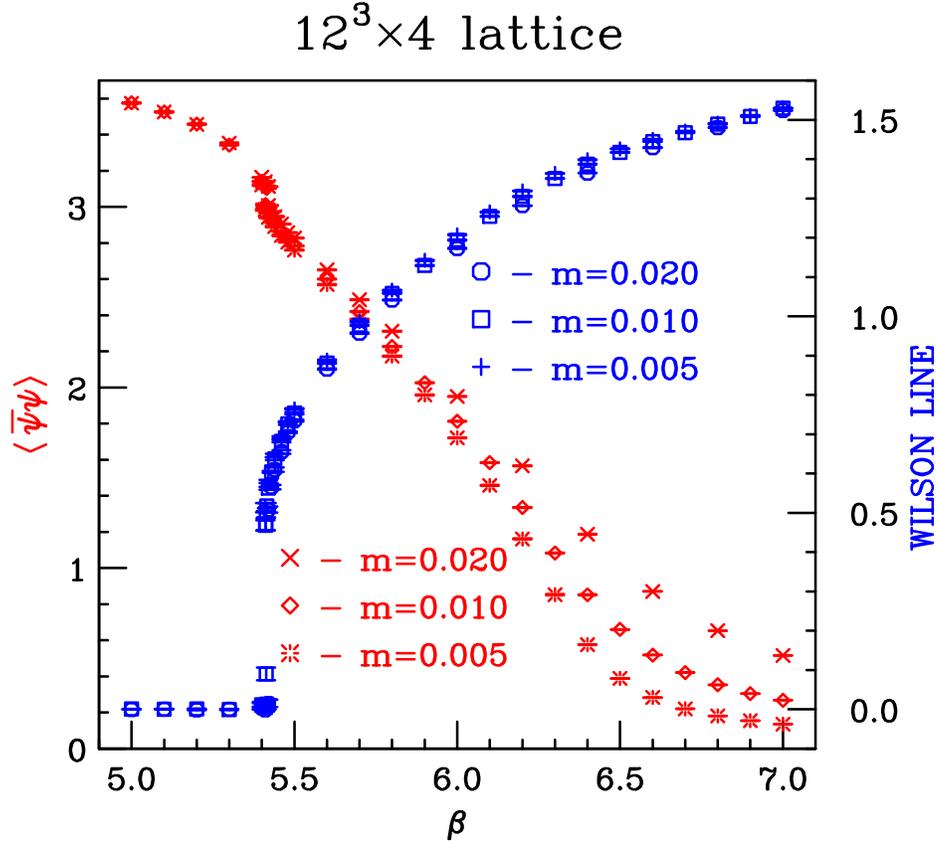}
\caption{Wilson line and $\langle\bar{\psi}\psi\rangle$
as functions of $\beta$ on a $12^3 \times 4$ lattice.}
\label{fig:wil-psi_12x4}
\end{figure}
Since the results from the 2 $N_t=4$ lattices are consistent, we present only
results from our $12^3 \times 4$ simulations. Figure~\ref{fig:wil-psi_12x4}
shows the colour-triplet Wilson Line(Polyakov Loop) and the chiral
condensate($\langle\bar{\psi}\psi\rangle$) as functions of
$\beta=6/g^2$, for each of the 3 quark masses on a $12^3 \times 4$
lattice. The deconfinement transition is marked by an abrupt increase in the
value of the Wilson Line. Chiral symmetry restoration occurs where the
chiral condensate vanishes in the chiral limit.

In contrast to what was found by DeGrand, Shamir and Svetitsky, we find well
separated deconfinement and chiral-symmetry restoration transitions. The
deconfinement transition occurs at $\beta=\beta_d$ where 
$\beta_d(m=0.005)=5.405(5)$, $\beta_d(m=0.01)=5.4115(5)$ and
$\beta_d(m=0.02)=5.420(5)$. The chiral transition, estimated from the peaks 
in the chiral susceptibility curves, occurs at $\beta_\chi=6.3(1)$.

Figure~\ref{fig:wil-psi_12x4} only accounts for the state with a real positive
Wilson Line in the deconfined regime. However, from the deconfinement 
transition up to $\beta \approx 5.9$ there exist long-lived states with the
Wilson Line oriented in the directions of the other 2 cube roots of unity.
However, these states are metastable, eventually decaying into the state with
a positive Wilson Line. Above $\beta \approx 5.9$, these complex Wilson Line
states disorder into a state with a negative Wilson Line.

\begin{figure}[htb]
\epsffile{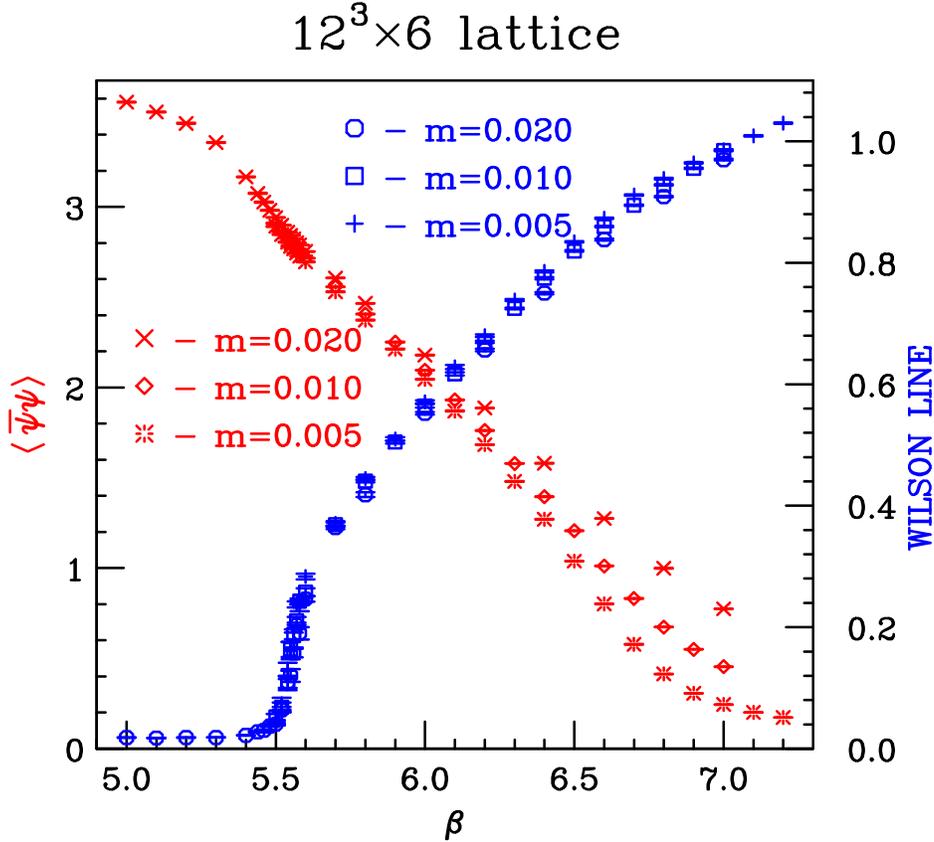}
\caption{Wilson Line and chiral condensate for the state with a real positive
Wilson Line as functions of $\beta$ for each of the 3 masses on a 
$12^3 \times 6$ lattice.}
\label{fig:rwil-psi_4x6}
\end{figure}
Let us now consider our $12^3 \times 6$ simulations. Again, the deconfinement
and chiral-symmetry restoring transitions are well-separated. Above the
deconfinement transition, we again find a clear 3-state signal. This time,
however, all 3 states appear equally stable. The system tunnels between these
3 states for the duration of the run until we are so far above the transition
that the relaxation time for tunneling exceeds the lengths of our runs. We
therefore artificially bin our `data' according to the phase of the Wilson
Line, into bins $(-\pi,-\pi/3)$, $(-\pi/3,\pi/3)$, $(\pi/3,\pi)$.

Figure~\ref{fig:rwil-psi_4x6} shows the Wilson Lines and chiral condensates
$\langle\bar{\psi}\psi\rangle$ for the central `positive' Wilson Line bin.
Figure~\ref{fig:cwil-psi_4x6} shows the Wilson Lines and chiral condensates
for the first and last `complex' and `negative' Wilson Line bins.
\begin{figure}[htb]
\epsffile{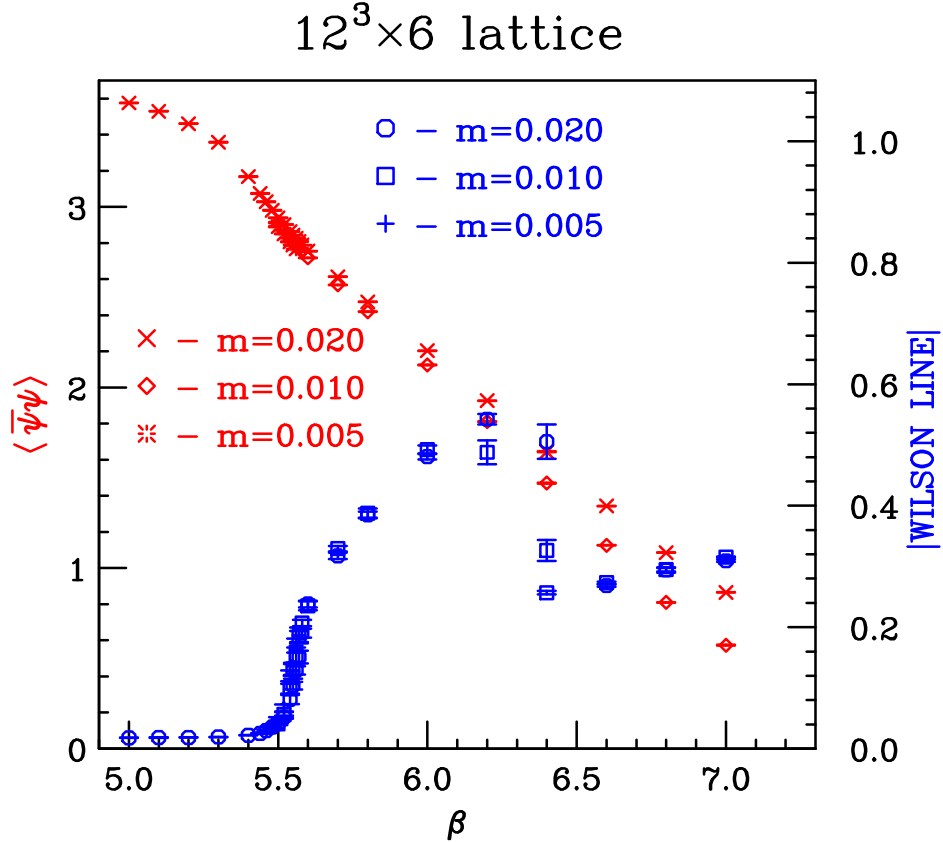}
\caption{Magnitude of the Wilson Line and chiral condensate for the state with 
a complex or negative Wilson Line as functions of $\beta$ for each of the 3 
masses on a $12^3 \times 6$ lattice.}
\label{fig:cwil-psi_4x6}                                         
\end{figure}             
The deconfinement transitions occur at $\beta_d(m=0.005)=5.545(5)$,
$\beta_d(m=0.01)=5.550(5)$ and $\beta_d(m=0.02)=5.560(5)$. Chiral-symmetry
restoration occurs at $\beta_\chi=6.6(1)$.

As for $N_t=4$, there is further transition between the deconfinement and
chiral transitions where the states with complex Wilson Lines disorder to
produce a state with a negative Wilson Line. This transition occurs at
$\beta \approx 6.4$ for $m=0.01$ and $\beta \approx 6.5$ for $m=0.02$.
This transition can be seen in fig.~\ref{fig:cwil-psi_4x6}.

\section{Discussion and conclusions}

We are studying the thermodynamics of Lattice QCD with 2 flavours of staggered
colour-sextet quarks. We find well separated deconfinement and chiral-symmetry
restoration transitions. This contrasts with the case of fundamental quarks,
where these 2 transitions are coincident, but is similar to the case of adjoint
quarks where again these 2 transitions are separate
\cite{Karsch:1998qj,Engels:2005te}.

We denote the value of $\beta=6/g^2$ at the deconfinement transition by 
$\beta_d$ and that at the chiral transition by $\beta_\chi$. In the chiral
limit $\beta_d \approx 5.40$ and $\beta_\chi=6.3(1)$ at $N_t=4$. At $N_t=6$
these values become $\beta_d \approx 5.54$ and $\beta_\chi=6.6(1)$. The
increase in the $\beta$s for both transitions from $N_t=4$ to $N_t=6$ is
consistent with their being finite temperature transitions for an
asymptotically free theory (rather than bulk transitions). If there is an IR
fixed point, we have yet to observe it. Our results suggest a Walking rather
than a conformal behaviour.

Why is this phase diagram so different from that for Wilson quarks (DeGrand,
Shamir and Svetitsky)? Is it because there is an infrared fixed point, and
we are on the strong-coupling side of it? Are our quark masses too large to
see the chiral limit? Is it because the flavour breaking of staggered quarks
does not allow a true chiral limit at fixed lattice spacing?

For the deconfined phase there is a 3-state signal, the remnant of now-broken
$Z_3$ symmetry. For $N_t=4$ the states with complex Polyakov Loops appear
metastable. For $N_t=6$ all 3 states appear stable. Breaking of $Z_3$ symmetry
is seen in the magnitudes of the Polyakov Loops for the real versus complex
states. Between the deconfinement and chiral transitions, we find a third
transition where the Wilson Lines in the directions of the 2 non-trivial roots
of unity change to real negative Wilson Lines. This transition occurs for
$\beta \approx 5.9$ ($N_t=4$) and $\beta \approx 6.4$--$6.5$ ($N_t=6$).
The existence of these extra states with Polyakov Loops which are not real
and positive has been predicted and observed by Machtey and Svetitsky.

Drawing conclusions from $N_t=4$ and $N_t=6$ is dangerous. We have
recently started simulations with $N_t=8$. We should also use smaller 
quark masses. At $N_t=6$, we need a second spatial lattice size.
To understand this theory more fully, we need to study its zero temperature
behaviour, measuring its spectrum, string tension, potential, $f_\pi$.... 
Measurement of the running of the coupling constant for weak coupling is needed.

We have recently started simulations with $N_f=3$, which is expected to be
conformal, to determine if it is qualitatively different from $N_f=2$.

\section*{Acknowledgements}

DKS is supported in part by the U.S. Department of Energy, Division of High
Energy Physics, Contract DE-AC02-06CH11357, and in part by the
Argonne/University of Chicago Joint Theory Institute. JBK is supported in part
by NSF grant NSF PHY03-04252. These simulations were performed on the Cray XT4,
Franklin at NERSC under an ERCAP allocation, and on the Cray XT5, Kraken at
NICS under an LRAC/TRAC allocation.

DKS thanks J.~Kuti, D.~Nogradi and F.~Sannino for helpful discussions.


\begin{thebibliography}{99}


\bibitem{Weinberg:1979bn}
  S.~Weinberg,
  Phys.\ Rev.\  D {\bf 19}, 1277 (1979).

\bibitem{Susskind:1978ms}
  L.~Susskind,
  Phys.\ Rev.\  D {\bf 20}, 2619 (1979).


\bibitem{Holdom:1981rm}
  B.~Holdom,
  Phys.\ Rev.\  D {\bf 24}, 1441 (1981).

\bibitem{Yamawaki:1985zg}
  K.~Yamawaki, M.~Bando and K.~i.~Matumoto,
  Phys.\ Rev.\ Lett.\  {\bf 56}, 1335 (1986).

\bibitem{Akiba:1985rr}
  T.~Akiba and T.~Yanagida,
  Phys.\ Lett.\  B {\bf 169}, 432 (1986).

\bibitem{Appelquist:1986an}
  T.~W.~Appelquist, D.~Karabali and L.~C.~R.~Wijewardhana,
  Phys.\ Rev.\ Lett.\  {\bf 57}, 957 (1986).









\bibitem{Kogut:1985pp}
  J.~B.~Kogut, J.~Polonyi, H.~W.~Wyld and D.~K.~Sinclair,
  Phys.\ Rev.\ Lett.\  {\bf 54}, 1475 (1985).

\bibitem{Fukugita:1987mb}
  M.~Fukugita, S.~Ohta and A.~Ukawa,
  Phys.\ Rev.\ Lett.\  {\bf 60}, 178 (1988).

\bibitem{Ohta:1991zi}
  S.~Ohta and S.~Kim,
  Phys.\ Rev.\  D {\bf 44}, 504 (1991).

\bibitem{Kim:1992pk}
  S.~y.~Kim and S.~Ohta,
  Phys.\ Rev.\  D {\bf 46}, 3607 (1992).

\bibitem{Brown:1992fz}
  F.~R.~Brown, H.~Chen, N.~H.~Christ, Z.~Dong, R.~D.~Mawhinney, W.~Schaffer and
  A.~Vaccarino,
  Phys.\ Rev.\  D {\bf 46}, 5655 (1992)
  [arXiv:hep-lat/9206001].

\bibitem{Iwasaki:1991mr}
  Y.~Iwasaki, K.~Kanaya, S.~Sakai and T.~Yoshie,
  Phys.\ Rev.\ Lett.\  {\bf 69}, 21 (1992).

\bibitem{Iwasaki:2003de}
  Y.~Iwasaki, K.~Kanaya, S.~Kaya, S.~Sakai and T.~Yoshie,
  Phys.\ Rev.\  D {\bf 69}, 014507 (2004)
  [arXiv:hep-lat/0309159].

\bibitem{Deuzeman:2008sc}
  A.~Deuzeman, M.~P.~Lombardo and E.~Pallante,
  Phys.\ Lett.\  B {\bf 670}, 41 (2008)
  [arXiv:0804.2905 [hep-lat]].

\bibitem{Deuzeman:2009mh}
  A.~Deuzeman, M.~P.~Lombardo and E.~Pallante,
  arXiv:0904.4662 [hep-ph].

\bibitem{Appelquist:2009ty}
  T.~Appelquist, G.~T.~Fleming and E.~T.~Neil,
  Phys.\ Rev.\  D {\bf 79}, 076010 (2009)
  [arXiv:0901.3766 [hep-ph]].

\bibitem{Appelquist:2007hu}
  T.~Appelquist, G.~T.~Fleming and E.~T.~Neil,
  Phys.\ Rev.\ Lett.\  {\bf 100}, 171607 (2008)
  [Erratum-ibid.\  {\bf 102}, 149902 (2009)]
  [arXiv:0712.0609 [hep-ph]].

\bibitem{Jin:2008rc}
  X.~Y.~Jin and R.~D.~Mawhinney,
  PoS {\bf LATTICE2008}, 059 (2008)
  [arXiv:0812.0413 [hep-lat]].

\bibitem{Jin:2009mc}
  X.~Y.~Jin and R.~D.~Mawhinney,
  PoS {\bf LAT2009}, 049 (2009)
  [arXiv:0910.3216 [hep-lat]].

\bibitem{Fodor:2009wk}
  Z.~Fodor, K.~Holland, J.~Kuti, D.~Nogradi and C.~Schroeder,
  arXiv:0907.4562 [hep-lat].

\bibitem{Fodor:2009ff}
  Z.~Fodor, K.~Holland, J.~Kuti, D.~Nogradi and C.~Schroeder,
  arXiv:0911.2463 [hep-lat].

\bibitem{Yamada:2009nt}
  N.~Yamada, M.~Hayakawa, K.~I.~Ishikawa, Y.~Osaki, S.~Takeda and S.~Uno,
  arXiv:0910.4218 [hep-lat].


\bibitem{Catterall:2007yx}
  S.~Catterall and F.~Sannino,
  Phys.\ Rev.\  D {\bf 76}, 034504 (2007)
  [arXiv:0705.1664 [hep-lat]].

\bibitem{Catterall:2008qk}
  S.~Catterall, J.~Giedt, F.~Sannino and J.~Schneible,
  JHEP {\bf 0811}, 009 (2008)
  [arXiv:0807.0792 [hep-lat]].

\bibitem{Catterall:2009sb}
  S.~Catterall, J.~Giedt, F.~Sannino and J.~Schneible,
  arXiv:0910.4387 [hep-lat].

\bibitem{DelDebbio:2008zf}
  L.~Del Debbio, A.~Patella and C.~Pica,
  arXiv:0805.2058 [hep-lat].

\bibitem{DelDebbio:2009fd}
  L.~Del Debbio, B.~Lucini, A.~Patella, C.~Pica and A.~Rago,
  arXiv:0907.3896 [hep-lat].

\bibitem{Bursa:2009we}
  F.~Bursa, L.~Del Debbio, L.~Keegan, C.~Pica and T.~Pickup,
  arXiv:0910.4535 [hep-ph].

\bibitem{Hietanen:2008mr}
  A.~J.~Hietanen, J.~Rantaharju, K.~Rummukainen and K.~Tuominen,
  JHEP {\bf 0905}, 025 (2009)
  [arXiv:0812.1467 [hep-lat]].

\bibitem{Hietanen:2009az}
  A.~J.~Hietanen, K.~Rummukainen and K.~Tuominen,
  arXiv:0904.0864 [hep-lat].


\bibitem{Shamir:2008pb}
  Y.~Shamir, B.~Svetitsky and T.~DeGrand,
  Phys.\ Rev.\  D {\bf 78}, 031502 (2008)
  [arXiv:0803.1707 [hep-lat]].

\bibitem{DeGrand:2009hu}
  T.~DeGrand,
  Phys.\ Rev.\  D {\bf 80}, 114507 (2009)
  [arXiv:0910.3072 [hep-lat]].

\bibitem{DeGrand:2008kx}
  T.~DeGrand, Y.~Shamir and B.~Svetitsky,
  Phys.\ Rev.\  D {\bf 79}, 034501 (2009)
  [arXiv:0812.1427 [hep-lat]].

\bibitem{Fodor:2008hm}
  Z.~Fodor, K.~Holland, J.~Kuti, D.~Nogradi and C.~Schroeder,
  arXiv:0809.4888 [hep-lat].


\bibitem{Machtey:2009wu}
  O.~Machtey and B.~Svetitsky,
  Phys.\ Rev.\  D {\bf 81}, 014501 (2010)
  [arXiv:0911.0886 [hep-lat]].


\bibitem{Kogut:2010cz}
  J.~B.~Kogut and D.~K.~Sinclair,
  arXiv:1002.2988 [hep-lat].


\bibitem{Karsch:1998qj}
  F.~Karsch and M.~Lutgemeier,
  Nucl.\ Phys.\  B {\bf 550}, 449 (1999)
  [arXiv:hep-lat/9812023].

\bibitem{Engels:2005te}
  J.~Engels, S.~Holtmann and T.~Schulze,
  Nucl.\ Phys.\  B {\bf 724}, 357 (2005)
  [arXiv:hep-lat/0505008].




\end{thebibliography}
\end{document}